\documentclass[amsmath,amssymb,aps,twocolumn,superscriptaddress,prl,10pt]{revtex4-2}

\usepackage{graphicx}
\graphicspath{{Figures/}}
\usepackage[normalem]{ulem}
\usepackage{MnSymbol}
\usepackage{hyperref}

\DeclareMathOperator{\Tr}{Tr}
\DeclareMathOperator{\const}{const}
\DeclareMathOperator{\var}{var}

\usepackage{titlesec}
\titleformat{\paragraph}[runin]{\normalfont\itshape\bfseries}{}{0pt}{}[---]
\titlespacing*{\paragraph}{\parindent}{0pt}{0pt}

\begin{document}

\title{Mesoscopic Fluctuations and Multifractality at and across Measurement-Induced Phase Transition}
\author{Igor Poboiko}
\author{Igor V. Gornyi}
\author{Alexander D. Mirlin}
\affiliation{\mbox{Institute for Quantum Materials and Technologies, Karlsruhe Institute of Technology, 76131 Karlsruhe, Germany}}
\affiliation{\mbox{Institut f\"ur Theorie der Kondensierten Materie, Karlsruhe Institute of Technology, 76131 Karlsruhe, Germany}}
\date{\today}

\begin{abstract}
We explore statistical fluctuations over the ensemble of quantum trajectories in a model of two-dimensional free fermions subject to projective monitoring of local charge across the measurement-induced phase transition. Our observables are the particle-number covariance between spatially separated regions, $G_{AB}$,
and the two-point density correlation function, $\mathcal{C}(r)$. Our results exhibit a remarkable analogy to Anderson localization, with $G_{AB}$ corresponding to two-terminal conductance and $\mathcal{C}(r)$ to two-point conductance, albeit with different replica limit and unconventional symmetry class, geometry, and boundary conditions.  In the delocalized phase, $G_{AB}$ exhibits ``universal'', nearly Gaussian, fluctuations with variance of order unity.  In the localized phase, we find a broad distribution of $G_{AB}$ with $\overline{-\ln G_{AB}} \sim L $ (where $L$ is the system size) and the variance $\var(\ln G_{AB}) \sim L^\mu$, and similarly for $\mathcal{C}(r)$, with $\mu \approx 0.5$. At the transition point, the distribution function of $G_{AB}$ becomes scale-invariant and $\mathcal{C}(r)$ exhibits multifractal statistics, $\overline{\mathcal{C}^{q}(r)}\sim r^{-q(d+1) - \Delta_{q}}$. We characterize the spectrum of multifractal dimensions $\Delta_q$. Our findings lay the groundwork for mesoscopic theory of monitored systems, paving the way for various extensions.

\end{abstract}

\maketitle
\paragraph{Introduction}
Many-body quantum systems subjected to quantum measurement exhibit remarkably rich physics. A peculiar property of quantum measurements is the non-unitary character of the associated dynamics of the quantum state, involving its collapse (strong or weak, depending on the measurement strength). The problem of the dynamics of a system under multiple measurements taking place randomly in space and time has been the subject of intense research in recent years. It was found that the competition between unitary dynamics and quantum measurements may result in transitions between phases with different scaling of the entanglement entropy $S$ (which is a measure of quantum information) with the size $\ell$ of a subsystem \cite{Li2018a, Skinner2019a, Chan2019a, Cao2019a, Szyniszewski2019a,  Li2019a, Bao2020a}, see also reviews \cite{Potter2022, Fisher2022}. 
Specifically, for frequent measurements, $S$ scales as the area of the subsystem boundary $\sim \ell^{d-1}$. In such an area-law phase, the entanglement is localized in the boundary region. When the measurement rate is lowered, the system may undergo a transition into a phase with a faster increase of $S(\ell)$, implying a delocalization of quantum information.

Many-body systems of free fermions, with local monitoring preserving Gaussianity of the quantum state \cite{Cao2019a, Turkeshi2021, Alberton2021a, Buchhold2021a, Coppola2022, Carollo2022, 
Szyniszewski2022, Poboiko2023a, Poboiko2023b, chahine2023entanglement, Lumia2023, starchl2024generalized, FavaNahum2024}, exhibit distinct physics. 
Specifically, it was found that, for $d=1$ complex fermions, the system is asymptotically ($L\to \infty$) in the area-law phase for any non-zero monitoring rate \cite{Poboiko2023a, FavaNahum2024}. For higher spatial dimensionality, $d > 1$, the system exhibits a transition between the area-law (``localized'') phase and a ``diffusive'' phase with  $\ell^{d-1} \ln \ell$ scaling of $S(\ell)$ 
\cite{Poboiko2023b, chahine2023entanglement}. These results are obtained analytically by mapping to a replica non-linear sigma model (NLSM), with its subsequent renormalization-group analysis, and supported by numerical simulations. The NLSM field theory (Appendix~\ref{sec:app:NLSM}) demonstrates a remarkable analogy between the measurement-induced transition in $d$ dimensions and Anderson localization transition in $d+1$ dimensions, albeit with important differences in the replica limit and the symmetry class. 

Dynamical evolution of a monitored system is characterized by a ``quantum trajectory'' determined by outcomes of all measurements, which are stochastic due to the nature of quantum measurements. Thus, the entanglement entropy or any other property of the system will, in fact, depend on a specific quantum trajectory. Most of the analytic and numerical studies deal with averaged observables---which is, in particular, sufficient to observe the phase transition. At the same time, fluctuations of observables over the ensemble of quantum trajectories are also of great interest. We will term these fluctuations ``mesoscopic'', in view of the above link to Anderson localization, where mesoscopic fluctuations are those over the ensemble of disorder realizations.
While fluctuations of some quantities appeared in several previous works on monitored systems, see, e.g., Refs.~\cite{Turkeshi2021, Zabalo2022, Jian2023, Doggen2023, chahine2023entanglement, LeGal2024entanglement, Paul2024, Fan2025,  Puetz2025}, their numerical and analytical investigations are still largely in their infancy.

In this Letter, we explore mesoscopic fluctuations of key observables across the measurement-induced transition in a free-fermion system. For numerical study, we use the $d=2$ model, in which the transition was established in Ref.~\cite{Poboiko2023b}.  It was found there that the particle-number covariance $G_{AB}$ plays a role of the scaling variable akin to the dimensionless conductance. Here, we show that $G_{AB}$ exhibits ``universal conductance fluctuations'' in the delocalized phase and another type of universality in the localized phase, where a broad distribution with the average $\overline{-\ln G_{AB}} \sim L $ and the variance $\var(\ln G_{AB}) \sim L^\mu$, with $\mu \approx 0.5$, sets in. The transition point is characterized by a scale-invariant distribution of $G_{AB}$. Furthermore, we study fluctuations of the density correlation function  $\mathcal{C}(r)$ (counterpart of two-point conductance). In particular,
we show that the critical point is characterized by multifractal statistics, $\overline{\mathcal{C}^{q}(r)}\sim r^{-q(d+1) - \Delta_{q}}$, and determine the spectrum of multifractal exponents $\Delta_q$. Our numerical results are in agreement with analytical findings obtained using the NLSM field theory.

\paragraph{Model}
We use the model of Ref.~\cite{Poboiko2023b}. 
The unitary part of the evolution is described by the nearest-neighbor tight-binding Hamiltonian 
$\hat{H}=-J\sum_{\left\langle \boldsymbol{x}\boldsymbol{x}^{\prime}\right\rangle }\left[\hat{\psi}_{\boldsymbol{x}}^{\dagger}\hat{\psi}_{\boldsymbol{x}^{\prime}}+\text{h.c.}\right]$,
defined on a $d$-dimensional cubic lattice with $L^d$ sites and periodic boundary conditions. The non-unitary part consists of stochastic projective measurements of local site occupancy numbers $\hat{n}_{\boldsymbol{x}}=\hat{\psi}_{\boldsymbol{x}}^{\dagger}\hat{\psi}_{\boldsymbol{x}}$ performed at each individual site with a rate $\gamma$. We focus on the case $d=2$.
At the same time, to address the transition analytically, we utilize the $\epsilon$-expansion for $d = 1+\epsilon$. 

The system is prepared in an arbitrary pure Gaussian (Slater-determinant) state at half-filling and is evolved for a sufficiently long time $T \to +\infty$ in order to reach a steady-state measure, which is independent of the initial state \cite{Poboiko2023a,Poboiko2023b}.
The final state is completely determined by a quantum trajectory, which includes positions, times, and outcomes of all measurements.
Throughout the evolution, the Gaussian property of the state is preserved, allowing its complete and efficient description via the correlation matrix ${\cal G}_{\boldsymbol{x}\boldsymbol{x}^{\prime}}\equiv\left\langle \hat{\psi}_{\boldsymbol{x}}^{\dagger}\hat{\psi}_{\boldsymbol{x}^{\prime}}\right\rangle $, where angular brackets denote quantum-mechanical average.

The key indicator of the measurement-induced transition in a charge-conserving free-fermion model is the particle-number covariance \cite{Poboiko2023b}, defined for two separated regions $A$ and $B$ as
\begin{equation}
G_{AB}\equiv\left\langle \hat{N}_{A}\right\rangle \left\langle \hat{N}_{B}\right\rangle -\left\langle \hat{N}_{A}\hat{N}_{B}\right\rangle 
= \sum_{\boldsymbol{x}\in A}
\sum_{\boldsymbol{x}^{\prime}\in B}
{\cal C}_{\boldsymbol{x}\boldsymbol{x}^{\prime}},
\label{eq:GAB-C}
\end{equation}
where ${\cal C}_{\boldsymbol{x}\boldsymbol{x}^{\prime}}$ is
the density correlation function,
\begin{equation}
{\cal C}_{\boldsymbol{x}\boldsymbol{x}^{\prime}}\equiv\left\langle \hat{n}_{\boldsymbol{x}}\right\rangle \left\langle \hat{n}_{\boldsymbol{x}^{\prime}}\right\rangle -\left\langle \hat{n}_{\boldsymbol{x}}\hat{n}_{\boldsymbol{x}^{\prime}}\right\rangle =\left|{\cal G}_{\boldsymbol{x}\boldsymbol{x}^{\prime}}\right|^{2}-{\cal G}_{\boldsymbol{x}\boldsymbol{x}}\delta_{\boldsymbol{x}\boldsymbol{x}^{\prime}}.
\label{eq:C}
\end{equation}

We note that the mutual information ${\cal I}(A:B)$, which is another observable commonly used to characterize measurement-induced transitions, is related to $G_{AB}$ via ${\cal I}(A:B)\approx(2\pi^{2}/3) \, G_{AB}$. 
Here, the sign $\approx$ indicates that an exact formula \cite{KlichLevitov} valid for any Gaussian state contains additional terms proportional to higher-order charge correlators. However, these terms are parametrically suppressed for $\gamma \ll J$ and, moreover, are numerically small \cite{Poboiko2023a} for any relation between $\gamma$ and $J$, so that this relation holds with very good accuracy. 

We consider a setup in which the linear sizes of regions $A$ and $B$ and the separation between them are of the order of the system size, $\ell_{A}, \ell_{B}, \ell_{AB} \sim L$. The averaged 
$G_{AB}$ 
is then a useful quantity to determine the position of the transition, in view of its distinct thermodynamic-limit behavior in the two phases and at criticality \cite{Poboiko2023b}:
\begin{equation}
\overline{G_{AB}} \sim\begin{cases}
gL^{d-1}, & \text{diffusive},\\
G_{c}=\const, & \text{critical},\\
\exp\left(-\ell_{AB}/\ell_{\text{loc}}\right), & \text{localized}.
\end{cases}
\label{eq:G-L-dependence}
\end{equation}
Here, the overbar denotes the averaging over quantum trajectories, and $\ell_{\text{loc}}$ is the localization length. 
This behavior is directly related 
[see Eq.~\eqref{eq:GAB-C}]
to the scaling of the average density correlation function:
\begin{equation}
\overline{{\cal C}_{\boldsymbol{x}\boldsymbol{x}^{\prime}}}
\sim
\begin{cases}
g|\boldsymbol{x}-\boldsymbol{x}^{\prime}|^{-(d+1)}, & \text{diffusive},\\
G_{c}|\boldsymbol{x}-\boldsymbol{x}^{\prime}|^{-2d}, & \text{critical},\\
\exp\left(-|\boldsymbol{x}-\boldsymbol{x}^{\prime}|/\ell_{\text{loc}}\right), & \text{localized}.
\end{cases}
\label{eq:C-scaling}
\end{equation}

In Ref.~\cite{Poboiko2023b}, we focused on averaged observables, demonstrated the measurement-induced phase transition, and determined the critical measurement rate $\gamma_c / J \approx 2.93$ for this model by using Eq.~\eqref{eq:G-L-dependence}.
Now, we turn to the analysis of mesoscopic fluctuations over the ensemble of quantum trajectories. Below, we set $J = 1$, so that the control parameter $\gamma / J$ becomes simply $\gamma$.

\paragraph{Numerical results}
We performed numerical simulations of the stochastic monitored evolution of a $d=2$ free-fermion system of size $L \times L$ with periodic boundary conditions (Appendix~\ref{sec:app:numerics}). For each individual quantum trajectory, we have extracted the distribution function of the particle-number covariance \eqref{eq:GAB-C} for regions $A$ and $B$ of size $L/4 \times L$ separated by distance $L/4$, and of the density correlation function \eqref{eq:C} at maximally separated points $\mathcal{C}_L \equiv \mathcal{C}_{\boldsymbol{x}\boldsymbol{x}^\prime}$ with $\boldsymbol{x}-\boldsymbol{x}^\prime = (L/2, L/2)$. Simulations were performed at the critical point, $\gamma = 2.93$, and in both diffusive ($\gamma = 0.5$ and $1.5$) and localized ($\gamma = 4.5$) phases, from $L = 12$ to $L = 44$.

\begin{figure}[ht]
  \centering
  \includegraphics[width=\columnwidth]{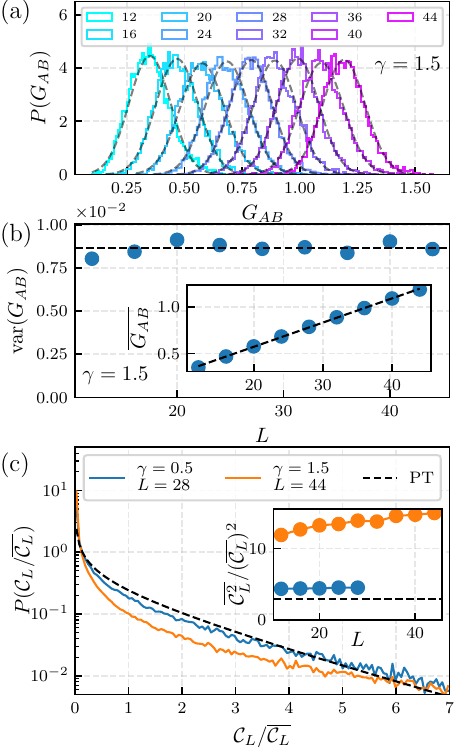}
  \caption{Statistical properties of the diffusive phase. (a) Distribution of $G_{AB}$ for $\gamma=1.5$ and different system sizes $L$. The distribution is nearly a perfect Gaussian, as demonstrated by Gaussian fits (dashed lines). (b) 
  System size (in-)dependence of variance of $G_{AB}$, demonstrating  ``universal conductance fluctuations'' with fitted value (dashed line) $\var(G_{AB}) = 8.64\cdot10^{-3}$. Inset: size-dependence of average $\overline{G_{AB}}$; dashed line: linear fit $G = (2.61~L + 4.94) \cdot 10^{-2}$, consistent with the large-$L$ scaling \eqref{eq:G-L-dependence}. (c) Distribution 
  $P(z = \mathcal{C}_L / \overline{{\cal C}_L})$ 
  of the density correlation function $\mathcal{C}_L$ normalized to its average for $\gamma=1.5$ and $0.5$; dashed line: Porter-Thomas (PT) distribution predicted analytically (see Appendix~\ref{sec:app:PT}) for $\gamma \ll 1$. Inset: $L$-dependence of the second moment $\overline{z^2}$; dashed line: the analytical prediction  $\overline{z^2}=3$ for $\gamma \ll 1$.}
  \label{fig:diffusive}
\end{figure}

We begin with our analysis of the diffusive phase.
Figure~\ref{fig:diffusive}(a) shows the $L$ dependence of the distribution function of the particle-number covariance. The distribution $P(G_{AB})$ is a nearly perfect Gaussian, down to the smallest system size.
Furthermore, the width of this distribution, shown on Fig.~\ref{fig:diffusive}(b), is size-independent, $\var(G_{AB}) \approx 8.61 \cdot 10^{-3}$. These results are in excellent agreement with the analytically predicted ``universal'' fluctuations (Appendix~\ref{sec:app:UCF}). 
Note that the ``universality'' here means independence of $L$ and $\gamma$ (within the diffusive phase).
At the same time, $\var(G_{AB})$ does depend on the chosen geometry of regions $A$ and $B$. In the inset of Fig.~\ref{fig:diffusive}(b), we check that the center of the distribution $P(G_{AB})$ shifts proportionally to $L$, in perfect agreement with the expected scaling \eqref{eq:G-L-dependence}.

In Fig.~\ref{fig:diffusive}(c), we show the distribution function 
$P(z = \mathcal{C}_L / \overline{{\cal C}_L})$
of the density correlation function $\mathcal{C}_L$ normalized to its average value $\overline{{\cal C}_L}$ for two values of $\gamma$ in the diffusive phase, $\gamma = 1.5$ and $0.5$. The distribution $P(z)$ and its moments $\overline{z^q}$ are $L$-independent (for $L$ larger than the correlation length $\ell_{\rm corr}$), as illustrated in the inset for $q=2$. Furthermore, when $\gamma$ is reduced, the distribution approaches the limiting form $P(z) =e^{-z/2}/\sqrt{2\pi z}$ (known as Porter-Thomas distribution in a different context), with $\overline{z^q} = (2q-1)!!$ in agreement with our analytical result (Appendix~\ref{sec:app:PT}). In the opposite case, when $\gamma$ approaches the transition point, the moments $\overline{z^q}$ diverge  $\sim \ell_{\rm corr}^{q-\Delta_q}$ due to multifractality at scales shorter than $\ell_{\rm corr}$ (see below the results for the multifractal spectrum $\Delta_q$ characterizing the critical point). The strong enhancement of $\overline{z^2}$ for $\gamma=1.5$ in comparison with its value $\overline{z^2}=3$ at $\gamma\ll 1$
seen in inset of Fig.~\ref{fig:diffusive}(c) is a manifestation of this behavior.

\begin{figure}[ht]
  \centering
  \includegraphics[width=\columnwidth]{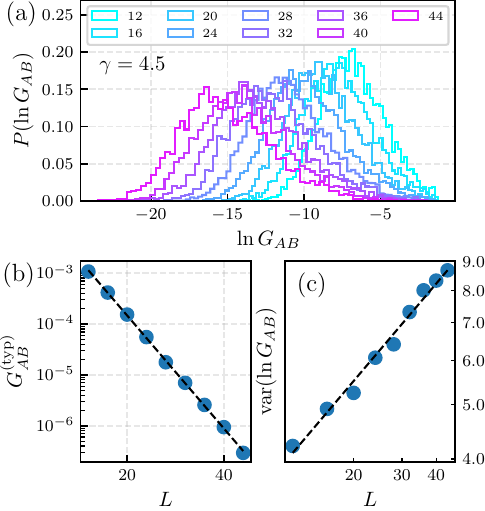}
  \caption{Statistics of $G_{AB}$ in the localized phase, $\gamma=4.5$. (a) Distribution function of the \emph{logarithm} of  $G_{AB}$ for different system sizes $L$. (b) $L$-dependence of \emph{typical} covariance $G_{AB}^{(\text{typ})}$, dashed line: exponential fit $G_{AB}^{\text{(typ)}} \sim \exp(-L / 4 \ell_{\text{loc}}^{\text{(typ)}})$ with $\ell_{\text{loc}}^{\text{(typ)}} \approx 0.98$. (c) $L$-dependence of the variance of $\ln G_{AB}$, dashed line: power-law fit $\var(\ln G_{AB})\sim L^{\mu}$ with $\mu \approx 0.58$.}
  \label{fig:localized}
\end{figure}

We turn now our attention to the other side of the transition and discuss the behavior in the localized phase. 
In Fig.~\ref{fig:localized}(a), we show the distribution of the \emph{logarithm} of the covariance $G_{AB}$. It is seen from Fig.~\ref{fig:localized}(b) that the center of the distribution moves linearly with $L$ in the negative direction, which corresponds to the typical value scaling as $G_{AB}^{\rm (typ)} = \exp\left( \overline{\ln G_{AB}}\right) \sim \exp(-L/4 \ell_{\rm loc}^{\rm (typ)})$, i.e., to exponential localization. Further, we observe that the distribution broadens with increasing $L$, which is quantified in Fig.~\ref{fig:localized}(c), where the variance of the distribution is shown as a function of $L$.  The data clearly indicate a power-law scaling $\var (\ln G_{AB}) \propto L^{\mu}$ with an exponent $\mu \approx 0.58$. 
We find a similar behavior for the distribution of 
$\ln {\cal C}_{\boldsymbol{x}\boldsymbol{x}^{\prime}}$, see  Supplemental Material (SM) \cite{SuppMat}, with a somewhat different numerical value of the exponent characterizing the scaling of the variance, $\mu \approx 0.43$. Asymptotically (at $L\to\infty$), the exponent $\mu$ should be the same for both observables, so that the apparent difference can be attributed to subleading (finite-size) corrections. This yields an estimate for the accuracy of the exponent, $\mu \approx 0.5 \pm 0.1$. 
The behavior that we find for the localized phase of the measurement problem is similar to that for the $D=d+1=3$ Anderson-localization problem that was shown to be related to the directed-polymer and Kardar-Parisi-Zhang (KPZ) problems in 2+1 dimensions, see Appendix~\ref{sec:app:KPZ}, for a summary of key results with references. It remains to be seen whether there is exact correspondence between the corresponding exponents. It is plausible (although remains to be proven) that our exponent $\mu$ is equal to $\mu = 2\delta \approx 0.48$ where $\delta$ is the KPZ growth exponent in 2+1 dimensions.  

\begin{figure}[ht]
  \centering
  \includegraphics[width=\columnwidth]{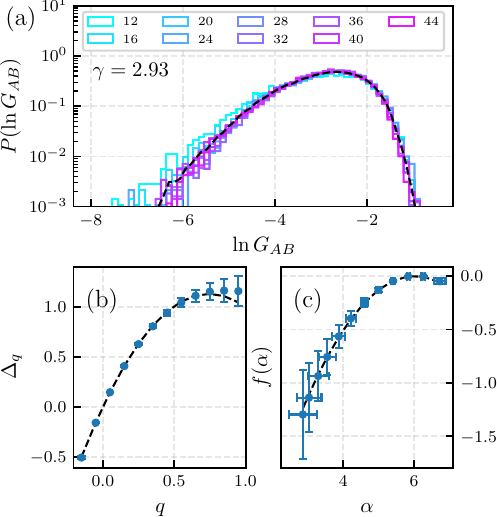}
  \caption{Properties of the critical point, $\gamma = 2.93$. (a) Scale-invariant distribution function of $ \ln G_{AB}$ for different system sizes $L$. (b) Anomalous dimensions $\Delta_q$ of moments of the density correlation function $\mathcal{C}_{\boldsymbol{x}\boldsymbol{x}^{\prime}}$. Dashed line is the best parabolic fit according to $\Delta_q = q + c q(1-q)$, with $c \approx 2.04$.   (c) Spectrum of multifractal dimensions $f(\alpha)$. Dashed line: 
  parabolic fit according to $f(\alpha) = -[\alpha-(4+c)]^2 / 4 c$, with $c \approx 2.03$.}
  \label{fig:critical}
\end{figure}

Finally, we analyze properties of systems at criticality, $\gamma \approx 2.93$. In Fig.~\ref{fig:critical}(a), we present the distribution function $P(\ln G_{AB})$. It is seen to be $L$ independent, in agreement with the expected scale invariance of the critical point. In Fig.~\ref{fig:critical}(b) and (c), we demonstrate another manifestation of criticality: the multifractal statistics of the correlation function $\mathcal{C}_{\boldsymbol{x}\boldsymbol{x}^{\prime}}$, which obeys the scaling $\overline{\mathcal{C}_{L}^{q}}\sim L^{-q(d+1)-\Delta_{q}}$. Here, the first term $-q(d+1)$ in the exponent corresponds to the diffusive scaling, while $\Delta_q$ is the anomalous dimension. The resulting exponents $\Delta_q$ are shown in Fig.~\ref{fig:critical}(b) (see SM \cite{SuppMat} for details of the fitting procedure). The non-linear dependence of $\Delta_q$ on $q$ is a manifestation of multifractality. 

We note that the \emph{average} correlation function 
$\overline{\mathcal{C}_{\boldsymbol{x}\boldsymbol{x}^{\prime}}}$
(corresponding to $q = 1$) 
is a correlation function of conserved currents and thus should follow at criticality the scaling \cite{Poboiko2023b}
$\overline{{\cal C}_{\boldsymbol{x}\boldsymbol{x}^{\prime}}}\sim|\boldsymbol{x}-\boldsymbol{x}^{\prime}|^{-2d}$, implying $\Delta_1 = d-1 = 1$, in good agreement with the numerical data. 
We further observe that the multifractality spectrum $\Delta_q$ is well approximated by the parabolic formula $\Delta_q = q + c q (1-q)$ shown by the dashed line in Fig.~\ref{fig:critical}(b). This formula corresponds to one-loop approximation, which is parametrically justified for a transition in $d=1+\epsilon$ dimensions with $\epsilon \ll 1$ (when the critical conductance is large), see Appendix~\ref{sec:app:multifractality}~\footnote{The $\epsilon$-expansion in one loop yields parabolic formula $\Delta_q = \epsilon q + \epsilon q (1-q)$, which for $\epsilon=1$ yields $c = 1$. This value is substantially smaller than $c \approx 2$ observed numerically, which is not unexpected since $\epsilon$-expansion is parametrically justified only for $\epsilon \ll 1$.}. Note that, while the parabolic approximation is found to hold with good accuracy also for $d=2$, it is only an approximation in view of higher-loop contributions to  $\Delta_q$.

An alternative (but equivalent) way to present the results on multifractality is as follows.  For each value of the correlation function, we introduce the 
exponent $\alpha=-\ln{\cal C}_{\boldsymbol{x}\boldsymbol{x}^{\prime}}/\ln\left|\boldsymbol{x}-\boldsymbol{x}^{\prime}\right|$. This quantity itself is random, and its distribution function acquires the form $P(\alpha)\sim L^{f(\alpha)}$, where $f(\alpha)$ is termed the singularity spectrum. It is then easy to show that $f(\alpha)$ is related to anomalous dimensions via the Legendre transformation:
\begin{equation}
f(\alpha)=q\Delta_{q}^{\prime}-\Delta_{q},\qquad
\alpha=
\Delta_{q}^{\prime} + (d+1).
\end{equation}
 The obtained function $f(\alpha)$ is shown in Fig.~\ref{fig:critical}(c). The position of its maximum, $\alpha_0 \approx 6$, determines the decay of the {\it typical} correlation function, ${\cal C}^{\rm (typ)}_{\boldsymbol{x}\boldsymbol{x}^{\prime}} \sim |\boldsymbol{x}-\boldsymbol{x}^{\prime}|^{-\alpha_0}$, which is considerably faster than the decay of the average, $\overline{{\cal C}_{\boldsymbol{x}\boldsymbol{x}^{\prime}}}\sim|\boldsymbol{x}-\boldsymbol{x}^{\prime}|^{-4}$.

\paragraph{Summary and outlook}
In this Letter, we have performed the analysis of mesoscopic fluctuations and multifractality across the measurement-induced transition in a monitored free-fermion system. 
Our numerical results for a $d=2$ model are in full consistency with analytical predictions. Our findings for the statistics of the particle-number covariance $G_{AB}$ and of the correlation function $\mathcal{C}(r)$ exhibit remarkable analogies with the mesoscopic physics of Anderson localization in disordered systems, with $G_{AB}$ and  $\mathcal{C}(r)$ being the counterparts of the two-terminal and two-point conductances, respectively. 

On the delocalized side (measurement rate $\gamma < \gamma_c$), we demonstrate the ``universal conductance fluctuations'' of  $G_{AB}$,  Fig.~\ref{fig:diffusive}(a,b). The moments of $\mathcal{C}(r)$ show a ``diffusive'' scaling, 
$\overline{\mathcal{C}^q}(r)\propto r^{-q(d+1)}$, with the distribution of $\mathcal{C}(r) / \overline{\mathcal{C}}(r)$ approaching the Porter-Thomas form for small $\gamma$ and manifesting multifractality at scales shorter than the correlation length when $\gamma$ increases towards the critical value $\gamma_c$.  In the localized phase ($\gamma > \gamma_c$), we find a broad distribution of $G_{AB}$, with $\overline{-\ln G_{AB}} \sim L $ and the variance $\var(\ln G_{AB}) \sim L^\mu$,  Fig.~\ref{fig:localized},
and similarly for 
$\mathcal{C}(r)$, with $\mu \approx 0.5$. It is plausible that, for both quantities, $\mu$ is given asymptotically by the KPZ theory in 2+1 dimensions, which implies $\mu\approx 0.48$.
At criticality ($\gamma=\gamma_c$), $G_{AB}$ develops a scale-invariant distribution and $\mathcal{C}(r)$ exhibits multifractality, Fig.~\ref{fig:critical}. The corresponding spectrum of  multifractal anomalous dimensions is well approximated by the parabolic (one-loop) form
$\Delta_q = q + c q (1-q)$ with $c \approx 2$. 

Concerning the link to the Anderson-localization problem in $d+1$ dimensions, it is worth emphasizing the following. First, the formulation of the measurement problem is very different from that of Anderson localization, and the connection made evident via the NLSM field theory, with the emergent isotropy of space-time, is rather non-trivial and remarkable. Second, there are important distinctions: the NLSM of the measurement problem is characterized by chiral symmetry, unconventional $R\to 1$ replica  limit, absorbing condition at the boundary of the $d+1$-dimensional system corresponding to the time at which the observables are studied and unconventional geometry of the ``terminals'' (in terminology of the transport problems) $A$ and $B$ that both belong to this boundary. 
These differences essentially affect many of key properties, including, in particular, the power laws in Eq.~\eqref{eq:C-scaling}, the multifractal spectrum $\Delta_q$ of the correlation function and its distribution in the diffusive phase, the shape of the distribution of $G_{AB}$ at criticality, as well as the numerical value of the variance of its ``universal'' fluctuations in the diffusive phase.

Our work, which largely lays the foundations of mesoscopic physics of monitored systems, paves a way for various extensions. This includes investigation of systems of various dimensionalities, as well as symmetry and topology classes. In connection with comments in the preceding paragraph, it will be very interesting to make a quantitative comparison with corresponding observables in the Anderson-localization problem.
Another interesting question is the effect of interaction on mesoscopic fluctuations and multifractality. 

\paragraph{Acknowledgments}
We thank I. Gruzberg and P. Ostrovsky for useful discussions and 
acknowledge support by the Deutsche Forschungsgemeinschaft (DFG, German Research Foundation) -- 553096561.

\bibliography{refs}

\appendix 

\clearpage

\setcounter{secnumdepth}{4}
\renewcommand{\theequation}{\Alph{subsection}\arabic{equation}}

\renewcommand{\thesubsection}{\Alph{subsection}}

\titleformat{\subsection}[runin]{\normalfont\itshape\bfseries}{Appendix~\thesubsection:~}{0pt}{}[---]
\titlespacing*{\subsection}{\parindent}{10pt}{0pt}

\renewcommand{\theparagraph}{\arabic{paragraph}}

\titleformat{\paragraph}[runin]{\normalfont\itshape\bfseries}{\theparagraph.~}{0pt}{}[---]

\section*{END MATTER}

\subsection{Outline of the analytical approach}

Here, we briefly outline the analytical approach. Detailed derivations with extensions will be published elsewhere \cite{TBP}.

\paragraph{Non-linear sigma model}
\label{sec:app:NLSM}
The system can be described by the $R \to 1$ limit of the NLSM of symmetry class BDI (chiral orthogonal) \cite{Poboiko2023a,Poboiko2023b,Poboiko2025}. It is a $(d+1)$-dimensional field theory, which describes fluctuations of the unitary matrix field $\hat{U}(\boldsymbol{r}) \in \mathrm{SU}(2R) / \mathrm{USp}(2R)$ in a semi-infinite space-time region where measurements are performed, $x^\mu = (v_0 t, \boldsymbol{x})$ with $t < 0$. Here $v_0$ is the root-mean-square group velocity 
$v_0 = \sqrt{2} J$.

The matrices $\hat{U}(\boldsymbol{r})$ 
belong to a product of the two-dimensional ``particle-hole'' space and $R$-dimensional replica space and can be parametrized as $\hat{U}={\cal V}\sigma^{y}{\cal V}^{T}\sigma^{y}$, with a matrix $\mathcal{V} \in \mathrm{SU}(2R)$ and $\sigma^y$ being the Pauli matrix acting on the particle-hole space. The action reads
\begin{equation}
S[\hat{U}]=\frac{g}{4}\int d^{d}\boldsymbol{x}\int_{-\infty}^{0}dx_0\Tr\left(\partial_{\mu}\hat{U}^{\dagger}\partial_{\mu}\hat{U}\right),
\label{eq:NLSM:action}
\end{equation}
and is supplied with the absorbing boundary conditions~\footnote{We note that \emph{average} two-point conductance at quantum Hall transition in a disordered system with absorbing boundary was studied in Ref.~\cite{Bettelheim2012quantum}} ${\hat{U}(\boldsymbol{x},x_0=0)=\hat{\mathbb{I}}}$. The bare value of the coupling constant $g$, playing the role of effective ``diffusion constant'', is related to microscopic parameters as $g = J / (2\sqrt{2} \gamma)$.

\paragraph{Diffusive phase: Fluctuations of the correlation function}
\label{sec:app:PT}
Utilizing the identity $\overline{{\cal C}_{\boldsymbol{x}\boldsymbol{x}^{\prime}}^{q}}=\overline{\left|{\cal G}_{\boldsymbol{x}\boldsymbol{x}^{\prime}}\right|^{2q}}$, we introduce $2q$ mutually different replica indices $r_1,\dots,r_q,r_1^\prime,\dots,r_q^\prime$ and obtain the following representation for the moments of the density correlation function:
\begin{equation}
\overline{{\cal C}_{\boldsymbol{x}\boldsymbol{x}^{\prime}}^{q}}=(-1)^q \left\langle {\cal O}^{(q)}_{r_{1}\dots r_{q},r_{1}^{\prime}\dots r_{q}^{\prime}}(\boldsymbol{x}){\cal O}^{(q)}_{r_{1}^{\prime}\dots r_{q}^{\prime},r_{1}\dots r_{q}}(\boldsymbol{x}^{\prime})\right\rangle,
\end{equation}
with operators
\begin{equation}
{\cal O}^{(q)}_{r_{1}\dots r_{q},r_{1}^{\prime}\dots r_{q}^{\prime}}(\boldsymbol{x})=g^q \det\nolimits^{1/2}\left(\left\{ \hat{{\cal J}}_{t,r_{i}r_{j}^{\prime}}(\boldsymbol{x},0)\right\} _{i,j=1}^{q}\right).
\label{eq:Cq:source}
\end{equation}
Here $\hat{{\cal J}}_t$ is the time component of the Noether current $\hat{{\cal J}}_{\mu}=-i\hat{U}^{\dagger}\partial_{\mu}\hat{U}$, and the determinant is taken for $2q\times2q$ sub-matrix corresponding to replica indices $r_i,r_j^\prime$. This generalizes the formula for the average correlation function \cite{Poboiko2023a} to its moments (requiring multiple replicas). In the semiclassical approximation governed by the parameter $g \gg 1$ or, equivalently, $\gamma \ll J$, the combinatorial factor arising from different contractions yields
\begin{equation}
\overline{{\cal C}_{\boldsymbol{x}\boldsymbol{x}^{\prime}}^{q}}=(2q-1)!!\,{\cal C}_{0}^{q}(\boldsymbol{x}-\boldsymbol{x}^{\prime}),
\end{equation}
implying a distribution of Porter-Thomas form,
\begin{equation}
P\left(z\equiv{\cal C}_{\boldsymbol{x}\boldsymbol{x}^{\prime}}/\overline{{\cal C}_{\boldsymbol{x}\boldsymbol{x}^{\prime}}}\right)=e^{-z/2}/\sqrt{2\pi z}.
\label{eq:C:PTDistribution}
\end{equation}
It is instructive to compare this result with the distribution of normalized transmission coefficients of waves propagating in random media \cite{Shapiro1986large,Kogan1993statistics,Nieuwenhuizen1995intensity,Kogan1995random-matrix,vanLangen1996nonperturbative,Mirlin1998intensity}, which has the form $P(z) = e^{-z}$ (Rayleigh statistics) in the weak-disorder limit. The primary reason for the difference between 
this formula and Eq.~\eqref{eq:C:PTDistribution} is a different symmetry class (BDI) relevant to our problem. 

\begin{figure}[hb]
    \centering
    \includegraphics[width=\columnwidth]{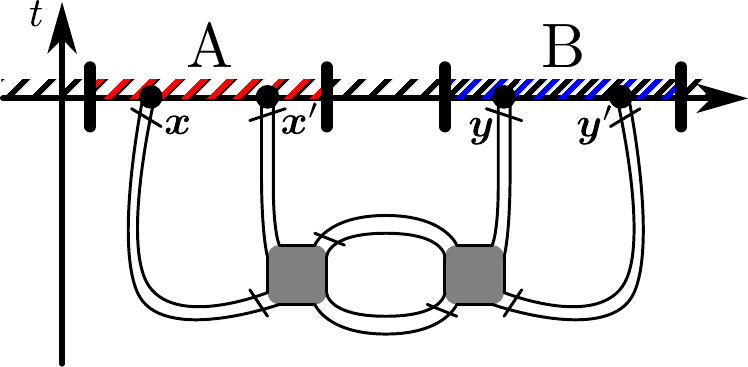}
    \caption{A diagram for mesoscopic variance of the particle-number covariance. Lines denote Gaussian propagators of \eqref{eq:NLSM:action}. Dashes on lines mark derivatives. Grey boxes are interaction vertices of NLSM modes. Additional diagrams arise from symmetrization with respect to $\boldsymbol{x}\leftrightarrow\boldsymbol{y}$ and $\boldsymbol{x}^\prime \leftrightarrow \boldsymbol{y}^\prime$ and from different ways to put two derivatives on legs of each interaction vertex.}
    \label{fig:ucf}
\end{figure}

\paragraph{Diffusive phase: Fluctuations of the particle-number covariance}
\label{sec:app:UCF}
The mesoscopic variance of the particle-number covariance $G_{AB}$ is obtained as
\begin{equation}
\var\left(G_{AB}\right)=\int_{A}d^{d}\boldsymbol{x}d^{d}\boldsymbol{x}^{\prime}\int_{B}d^{d}\boldsymbol{y}d^{d}\boldsymbol{y}^{\prime}{\cal C}^{(2)}(\boldsymbol{x},\boldsymbol{y},\boldsymbol{x}^{\prime},\boldsymbol{y}^{\prime}),
\end{equation}
with
\begin{multline}
{\cal C}^{(2)}(\boldsymbol{x},\boldsymbol{y},\boldsymbol{x}^{\prime},\boldsymbol{y}^{\prime})\equiv\overline{{\cal C}_{\boldsymbol{x}\boldsymbol{y}}{\cal C}_{\boldsymbol{x}^{\prime}\boldsymbol{y}^{\prime}}}-(\overline{{\cal C}_{\boldsymbol{x}\boldsymbol{y}}})(\overline{{\cal C}_{\boldsymbol{x}^{\prime}\boldsymbol{y}^{\prime}}})\\
=g^4\Big\llangle {\cal J}_{t,r_{1}r_{1}^{\prime}}(\boldsymbol{x},0){\cal J}_{t,r_{1}^{\prime}r_{1}}(\boldsymbol{y},0)\\\
\times{\cal J}_{t,r_{2}r_{2}^{\prime}}(\boldsymbol{x}^{\prime},0){\cal J}_{t,r_{2}^{\prime}r_{2}}(\boldsymbol{y}^{\prime},0)\Big\rrangle,
\end{multline}
where $r_{1,2},r_{1,2}^\prime$ are mutually different replica indices and $\llangle \ldots \rrangle$ denote the connected NLSM correlation function. Such a correlation function is zero in the semiclassical approximation. The first-order correction with respect to the non-linear interaction vertices also vanishes due to the replica structure. Thus, the main contribution in the limit $\gamma \ll J$ arises from the second order of perturbation theory, as shown in Fig.~\ref{fig:ucf}.

The exact expression for the diagram in 
Fig.~\ref{fig:ucf}, together with similar diagrams arising from symmetrization (see figure caption), 
 leads to a cumbersome multidimensional integral. For our purposes, it will be sufficient to estimate its dependence on $g$ and $L$ (without calculating the numerical factor), in a geometry where all length scales are of order $L$.  Four external vertices, together with integration over their position, yield a factor $\sim g^4 L^{4d}$. Two interaction vertices, with integration over their position, yield a factor $\sim g^2 L^{2(d+1)}$. Six Green's function yield factor $\sim g^{-6} L^{-6(d-1)}$. Finally, eight gradients yield a factor $\sim L^{-8}$. Combining everything, we find that the result is proportional to $\propto g^0 L^0$. Thus, $\var(G_{AB})$ is given by a \emph{universal number}, which does not depend on $\gamma$ and on  $L$. (Of course, it depends on the chosen geometry of regions $A$ and $B$.)

We also note that higher cumulants arise only in the higher orders of perturbation theory, which is controlled by a large parameter $G \sim g L^{d-1} \gg 1$, and thus are suppressed in the diffusive phase. Thus, we conclude that the distribution function of $G_{AB}$ in the diffusive phase is Gaussian to the leading order.

Similarity to ``universal conductance fluctuations'' studied in the context of mesoscopic physics of disordered metals, see Refs.~\cite{LeeStone1985,Lee1987universal,Kane1988,Altshuler1985}, should be emphasized. This analogy is truly remarkable, taking into account differences in original formulation of the problem and in observables, as well as in symmetry class, boundary conditions, and the replica limit of the NLSM (with disordered systems corresponding to the replica limit $R \to 0$).

\paragraph{Multifractality of the correlation function at the transition point}
\label{sec:app:multifractality}

\begin{figure}[hb]
    \centering
    \includegraphics[width=\columnwidth]{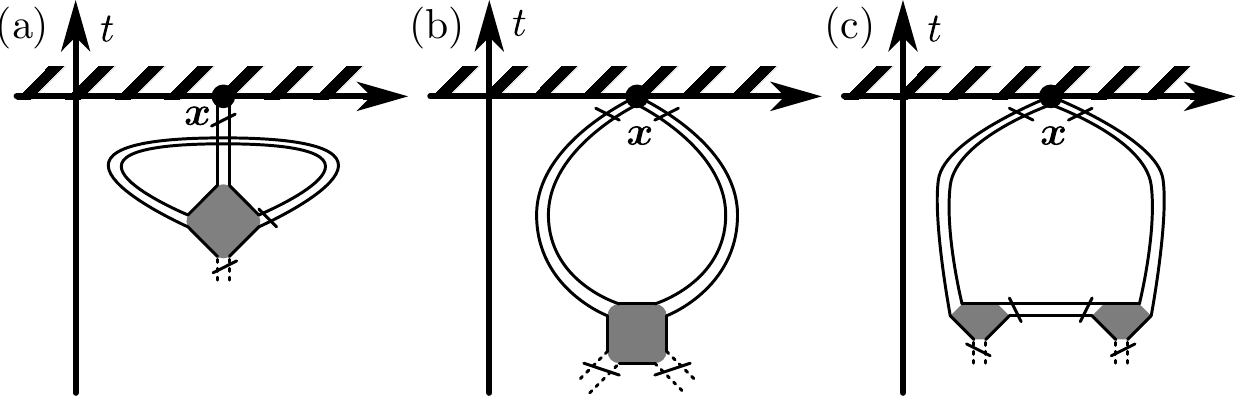}
    \caption{Three diagrams contributing to renormalization of operator $\mathcal{O}^{(q)}(\boldsymbol{x})$, Eq.~\eqref{eq:Cq:source}, in one-loop order. For diagram (a), which 
    appears already for  $q=1$
    \cite{Poboiko2023a,Poboiko2023b}), only a single leg from $\mathcal{O}^{(q)}(\boldsymbol{x})$
    is contracted with the interaction vertex, while remaining $q-1$ legs are not affected (and not shown).
    For diagrams (b) and (c), two legs from $\mathcal{O}^{(q)}(\boldsymbol{x})$ are contracted with interaction vertices, while the remaining $q-2$ legs are not involved.  Bold lines denote propagators of fast modes, while external dashed lines correspond to the slow fields.
    }
    \label{fig:multifractality}
\end{figure}

The multifractality is revealed by scaling of the moments $\overline{{\cal C}^{q}(\boldsymbol{x},\boldsymbol{x}^{\prime})}$ related to anomalous dimensions of the operators $\mathcal{O}^{(q)}(\boldsymbol{x})$,  Eq.~\eqref{eq:Cq:source}. In the background-field formalism, there are three diagrams contributing to the renormalization of this operator, see Fig.~\ref{fig:multifractality}. For clarity, we first consider the case $q = 2$ and then restore the $q$-dependence.

For $q=2$, the operator $\mathcal{O}^{(2)}$ is not an eigenoperator of the renormalization group (RG) but has a non-zero overlap with the most relevant eigenoperator  $g^2 \Tr(\mathcal{J}_t^2(\boldsymbol{x}))$. The corresponding RG flow reads:
\begin{equation}
\frac{d\ln Z_{2}}{d\ln\ell}=\frac{R}{2\pi G} + O(G^0).
\end{equation}
In the $\epsilon = (d-1)$-expansion, the critical value of the conductance is given by $G_c = R / 2 \pi \epsilon$ \cite{Poboiko2023b,Poboiko2025}, and thus at the critical point one has $Z_{2}(\ell)\sim\left(\ell/\ell_{0}\right)^{\epsilon}$. The renormalized second moment of the correlation function is then determined by:
\begin{equation}
\overline{{\cal C}^{2}(\ell)}\sim{g^{2}(\ell)Z_{2}^{2}(\ell)}
\ell^{-2(d+1)}
\sim\ell^{-2(d+1)},
\end{equation}
implying $\Delta_2 = 0$ to the one-loop order, i.e., $\Delta_2 = O(\epsilon^2)$. 

To restore the dependence on $q$, we note that (i) for combinatorial reasoning, $\Delta_q$ in the one-loop approximation has to be a second-degree polynomial in $q$, and (ii) the values $\Delta_0 = 0$ and $\Delta_1 = d-1$ are fixed. This yields
\begin{equation}
\Delta_{q}=\epsilon q+\epsilon q(1-q)+O(\epsilon^{2}).
\end{equation}
Here, the first term originates from the diagram (a) in Fig.~\ref{fig:multifractality} and the second term from diagrams (b) and (c), with factors $q$ and $q(q-1)/2$ counting possibilities to choose one leg and a pair of legs, respectively. 

\vspace*{0.2cm}

\subsection{Mesoscopic fluctuations for strong Anderson localization}
\label{sec:app:KPZ}

\vspace*{-0.35cm}

Here we recall key analytical and numerical results for the distribution of conductance $G$ (and related observables governed by tails of localized wave functions) in the strong-localization regime for the problem of Anderson localization in $D = d+1$ dimensions. 
The average value $\overline{-\ln G}$ grows linearly with $L$ for any $D$, and $\text{var}(\ln G)$ scales as $L^\mu$. For $D = 1$ (including also quasi-one-dimensional geometry), the problem is exactly solvable and $\mu=1$ \cite{mirlin00}. For $D > 1$, it was argued that the forward-scattering approximation contains all the essential physics, implying that the exponent $\mu/2$ is given by the droplet exponent $\theta$ of the directed polymer problem---or, equivalently, by the growth exponent $\delta$ of the KPZ problem---in $d+1$ dimensions \cite{Medina1992quantum,Pietracaprina2016forward,Mu2024}. For $D=2$, this exponent is known exactly, yielding $\mu=2/3$. This value has been confirmed by numerical simulations of 
strong Anderson localization in two-dimensional systems \cite{Prior2005conductance,Somoza2007universal,Lemarie2019glassy,Mu2024,swain2025_2d_anderson}. For $D=3$, corresponding to spatial dimension $d = 2$ studied in the present paper, intensive numerical simulations of the KPZ problem yielded $\delta \simeq 0.24$ \cite{Forrest1990surface,Kelling2011extremely,Monthus2012random}, suggesting $\mu \simeq 0.48$ for Anderson localization. Numerical simulations of $D=3$ Anderson localization \cite{Somoza2006conductance} yielded a close but somewhat smaller value, $\mu \approx 0.40$. At the same time, 
Ref.~\cite{Pietracaprina2016forward} used the forward-scattering approximation and obtained a value $\mu \approx 0.56$, slightly above the KPZ (directed-polymer) value.

\subsection{{Numerical simulation details}}
\label{sec:app:numerics}

The numerical analysis was performed via direct exact simulations of the evolution of the full correlation matrix $\mathcal{G}_{\boldsymbol{x}\boldsymbol{x}^\prime}$. To explore the scale dependence of observables, the system size $L$ was varied between $L = 12$ and $L = 44$ in steps of 4. To gather sufficient statistics, we have sampled 1000 individual trajectory realizations for each system size. In addition, we have utilized spatial averaging over different positions in the system to further increase the statistical ensemble. Further details of simulations can be found in SM \cite{SuppMat}.

\vspace{1cm}

\widetext
\clearpage

\setcounter{secnumdepth}{1}
\renewcommand{\theequation}{S\arabic{equation}}
\renewcommand{\thefigure}{S\arabic{figure}}
\renewcommand{\thesection}{S\arabic{section}}
\setcounter{equation}{0}
\setcounter{figure}{0}

\begin{center}
    \Large Supplemental Materials to ``Mesoscopic Fluctuations and Multifractality at and across Measurement-Induced Phase Transition''
\end{center}

\section{Numerical study of multifractality of the correlation function \texorpdfstring{$\mathcal{C}_{\boldsymbol{x}\boldsymbol{x}^\prime}$}{C(x,x')} at criticality}

\begin{figure}[ht]
\centering
\includegraphics[width=0.2\columnwidth]{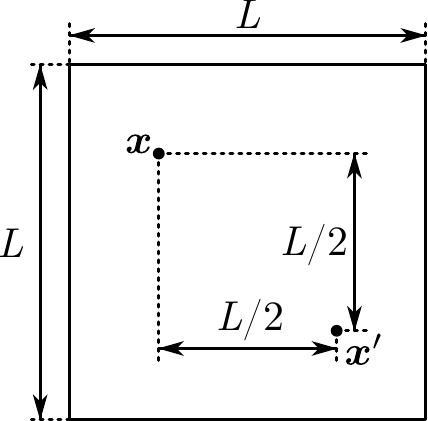}
\caption{Schematic representation of geometry, in which the correlation function $\mathcal{C}_L = \mathcal{C}_{\boldsymbol{x}\boldsymbol{x}^\prime}$ was considered in the main text, and, in particular, in which the multifractality was analyzed. The system has size $L \times L$ and has periodic boundary conditions; points $\boldsymbol{x}$ and $\boldsymbol{x}^\prime$ are taken at maximal separation $\boldsymbol{x}-\boldsymbol{x}^\prime = (L/2, L/2)$.}
\label{fig:C-scheme}
\end{figure}

To analyze numerically the multifractal behavior of the density correlation function at the critical point, for each system size $L$, we have gathered statistics of $\mathcal{C}_{L} \equiv \mathcal{C}_{\boldsymbol{x}\boldsymbol{x}^\prime}$, for maximally separated values $\boldsymbol{x}-\boldsymbol{x}^\prime = (L/2, L/2)$, see Fig.~\ref{fig:C-scheme}. To determine the anomalous exponents $\Delta_q$ governing the power-law scaling at criticality,
\begin{equation}
\overline{ {\cal C}_{L}^{q} } \sim L^{-\left[q(d+1)+\Delta_{q}\right]},
\end{equation}
we have performed linear fits of the form:
\begin{equation}
\label{eq:tau-regression}
\ln\left(\overline{ \mathcal{C}_{L}^{q} }\right) =a_q-\left[q(d+1)+\Delta_{q}\right]\ln L,    
\end{equation}
with $a_q$ and $\Delta_q$ being fitting parameters.

Furthermore, differentiating Eq.~\eqref{eq:tau-regression} with respect to $q$ yields
\begin{equation}
\label{eq:alpha-regression}
\frac{\overline{ {\cal C}_{L}^{q}\ln{\cal C}_{L} } }{\overline{ {\cal C}_{L}^{q} } }=b_{q}-\alpha_{q}\ln L,
\end{equation}
with $b_q = a^\prime_q$. This equation was used to extract values $\alpha_q$ from another linear fitting procedure, treating $b_q$ and $\alpha_q = d+1 + \Delta^\prime_q$ as fitting parameters.


\begin{figure}[ht]
\centering
\includegraphics[width=0.6\columnwidth]{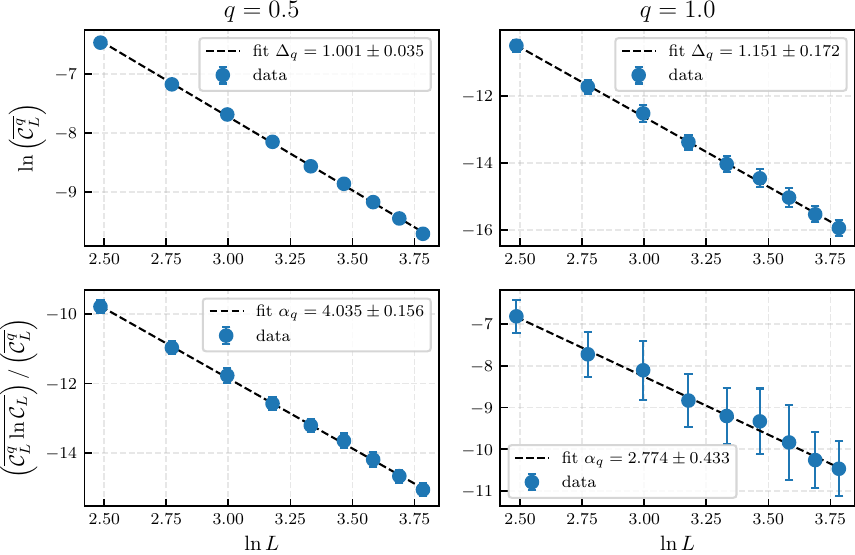}
\caption{Examples of fits according to Eqs.~(\ref{eq:tau-regression}) and (\ref{eq:alpha-regression}) (top and bottom panels, correspondingly) for $q = 0.5$ (left panels) and $q = 1.0$ (right panels).}
\label{fig:fits}
\end{figure}


The fitting procedure took into account statistical error bars of the left-hand side of Eqs.~(\ref{eq:tau-regression},\ref{eq:alpha-regression}).
To estimate these error bars, the random subsampling method (also known as ``$m$-out-of-$n$ bootstrap'') was used \cite{Subsampling1999}. Out of $n$ available values of $\mathcal{C}_{L}$ for each system size $L$, we have sampled $n_{\text{samples}} = 4096$ subsamples of smaller size $m = n^\gamma$; for practical purposes, the value $\gamma = 0.7$ was chosen. The left-hand side was then calculated for each subsample, yielding the approximated distribution of the corresponding value. The latter was then used to estimate the error bar for each data point. Examples of power-law fits for two values $q = 0.5$ and 1.0 are presented in Fig.~\ref{fig:fits}. Note that the obtained value $\Delta_1 \approx 1.151 \pm 0.172$ is consistent with the analytical prediction of exact value $\Delta_1 = 1$.

\section{Distribution of \texorpdfstring{$\mathcal{C}_{\boldsymbol{x}\boldsymbol{x}^\prime}$}{C(x,x')} in the localized phase and comparison to the distribution of \texorpdfstring{$G_{AB}$}{G(A,B)}}

\begin{figure}[ht]
\centering
\includegraphics[width=0.4\textwidth]{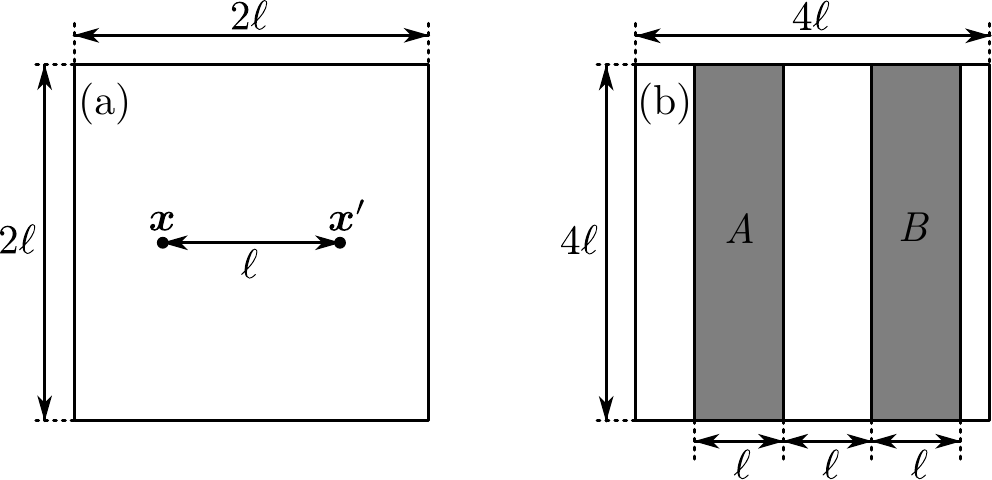}
\caption{Geometry, in which the mesoscopic fluctuations of correlation function $\mathcal{C}_{\boldsymbol{x}\boldsymbol{x}^\prime}$ and particle-number covariance $G_{AB}$ were analyzed in the localized phase. (a) The correlation function $\mathcal{C}_{\boldsymbol{x}\boldsymbol{x}^\prime}$ was analyzed in a system of size $2\ell \times 2\ell$ with periodic boundary conditions, with points separated by $\boldsymbol{x}-\boldsymbol{x}^\prime = (\ell, 0)$ or $(0, \ell)$. (b) The particle number covariance $G_{AB}$ was analyzed in system of size $4\ell \times 4\ell$ with periodic boundary conditions, with regions $A$ and $B$ of size $\ell \times 4\ell$ separated by distance $\ell$.
(The same geometry, with $4\ell= L$, was used for a study of the distribution of $G_{AB}$ in the diffusive phase and at criticality in the main text.)
}
\label{fig:scheme:insulator}
\end{figure}

\begin{figure}[ht]
\centering
\includegraphics[width=\textwidth]{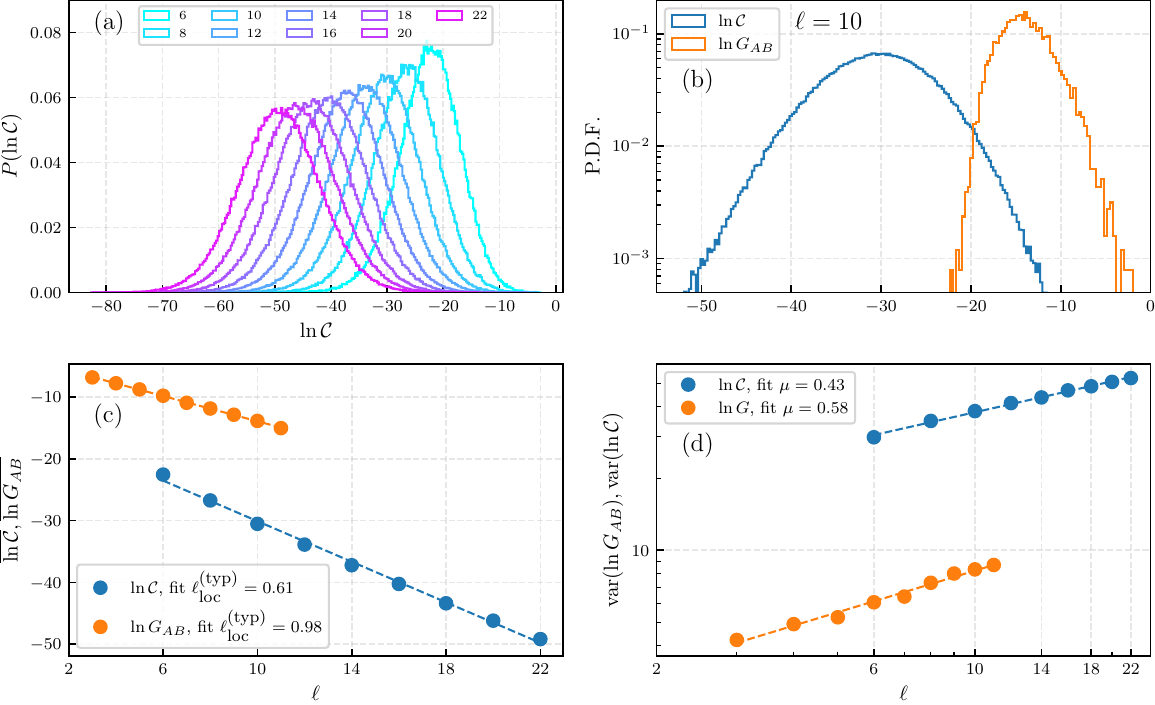}
\caption{Comparison between statistical properties of $\ln \mathcal{C}_{\boldsymbol{x}\boldsymbol{x}^\prime}$ and $\ln G_{AB}$ in the localized phase, $\gamma=4.5$, with $\ell$ being the distance in both cases. The correlation function $\mathcal{C}_{\boldsymbol{x}\boldsymbol{x}^\prime}$ was evaluated for $\boldsymbol{x} - \boldsymbol{x}^\prime = (\ell, 0)$ in a system of size $2\ell \times 2\ell$ (Fig.~\ref{fig:scheme:insulator}(a)) whereas the particle-number covariance $G_{AB}$ was calculated in a system of size $4\ell \times 4\ell$ with geometry of subsystems $A$ and $B$ as described in the main text. Specifically, the subsystems $A$ and $B$ are of size $4\ell \times \ell$ and are separated by a distance $\ell_{AB} = \ell$ (Fig.~\ref{fig:scheme:insulator}(b)), thus allowing a direct comparison of both distributions. (a) Probability distribution function of $\ln \mathcal{C}$ for different values of $\ell$ (shown in the legend). (b) Comparison of distributions of $\ln \mathcal{C}$ and $\ln G_{AB}$. (c) Fits for typical localization length from the behavior $\overline{\ln \mathcal{C}} = \const -\ell / \ell_{\text{loc}}^{\text{(typ)}}$ and $\overline{\ln G_{AB}} = \const - \ell / \ell_{\text{loc}}^{\text{(typ)}}$.
(d) Fits for scaling of the variance $\var (\ln \mathcal{C}) \sim \ell^{\mu}$ and $\var (\ln G_{AB}) \sim \ell^\mu$. In the asymptotic limit $\ell \to \infty$, the values of $\ell_{\text{loc}}^{\text{(typ)}}$ and $\mu$ [panels (c) and (d)]
should be the same for both distributions, so that differences in obtained numerical values are attributed to finite-size effects.}
\label{fig:PDF:C:localized}
\end{figure}

An analysis similar to the one presented in the main text (see Fig.~\ref{fig:localized})
for the distribution of the covariance $G_{AB}$ in the localized phase has also been performed for the density correlation function $\mathcal{C}_{\boldsymbol{x}\boldsymbol{x}^\prime}$. To simplify the comparison between the two distributions, we have considered here $\mathcal{C}_{\boldsymbol{x}\boldsymbol{x}^\prime}$ with $\boldsymbol{x} -\boldsymbol{x}^\prime = (L/2, 0)$ or $(0, L/2)$, see Fig.~\ref{fig:scheme:insulator}(a). The distribution function of 
$\ln \mathcal{C}_{\boldsymbol{x}\boldsymbol{x}^\prime}$ for this choice of $\boldsymbol{x} -\boldsymbol{x}^\prime$ is presented in Fig.~\ref{fig:PDF:C:localized}(a) for $L$ from 12 to 44 (i.e., for the distance $\ell = |\boldsymbol{x} -\boldsymbol{x}^\prime|$ from 6 to 22). It exhibits the same features as the distribution of $\ln G_{AB}$ whose evolution with $L$ is shown in Fig.~\ref{fig:localized}(a) of the main text.
Specifically, upon increasing the system size, the distribution $P(\ln \mathcal{C})$ shifts towards smaller values and broadens. The shift of the position of the maximum of the distribution is approximately linear with $\ell$, which corresponds to an exponential decay of the typical value $\mathcal{C}^{\mathrm{(typ)}} \sim \exp(-\ell /  \ell_{\text{loc}}^{\text{(typ)}})$.


To better visualize the similarity and difference between the two distributions, we compare in Fig.~\ref{fig:PDF:C:localized}(b) both  distribution for the same distance $\ell=10$ between the relevant subsystems
(points 
$\boldsymbol{x}$ and $\boldsymbol{x}^\prime$
for
$\mathcal{C}_{\boldsymbol{x}\boldsymbol{x}^\prime}$ and regions $A$ and $B$ for $G_{AB}$, see Fig.~\ref{fig:scheme:insulator}). It is seen that the distribution of $\ln G_{AB}$ is shifted towards smaller (by absolute value) values and is narrower in comparison with the distribution of $\ln\mathcal{C}$. In panels (c) and (d), we present the averages and the variances of both distributions. The $\ell$ dependences of average logarithms are approximately linear as expected, implying an exponential localization $\sim 
\exp(- \ell / \ell_{\text{loc}}^{\text{(typ)}})$ of typical values.
The fits yield estimates of the typical localization length 
[see Fig.~\ref{fig:PDF:C:localized}(c)]
$\ell_{\text{loc}}^{\text{(typ)}} \approx 0.61$ for $\overline{\ln \mathcal{C}}$ and $\ell_{\text{loc}}^{\text{(typ)}} \approx 0.98$ for 
$\overline{\ln G_{AB}}$ (the latter value is also given in the main text). Asymptotically (i.e., at $\ell \to \infty$), the localization length should be the same for both observables (as discussed in more detail below). The difference is due to finite-size effects associated with subleading corrections to scaling. Further, the broadening of the distributions 
[Fig.~\ref{fig:PDF:C:localized}(d)]
follows power-law scaling with the system size: $\var(\ln \mathcal{C}) \sim L^{\mu}$ with $\mu \approx 0.43$ and (as also shown in the main text) $\var (\ln G_{AB}) \sim L^{\mu}$ with $\mu \approx 0.58$. Again, in the asymptotic limit $\ell \to \infty$, the values of $\mu$ should be the same for both observables, and the difference is attributed to finite-size effects.

We explain now why the values of the typical localization length $\ell_{\text{loc}}^{\text{(typ)}}$, and also of the exponent $\mu$, should be asymptotically the same for both distributions and clarify the source of finite-size deviations. The particle-number covariance is related to the correlation function as 
\begin{equation}
\label{eq:supp:GAB-C}
G_{AB}=\sum_{\boldsymbol{x}\in A}\sum_{\boldsymbol{x}^{\prime}\in B}{\cal C}_{\boldsymbol{x}\boldsymbol{x}^{\prime}}.
\end{equation}
For large $\ell$, we have in the localized phase
\begin{equation}
\label{eq:average-ln-C}
\overline{\ln \mathcal{C}} \simeq  - \ell / \ell_{\text{loc}}^{\text{(typ)}} \end{equation} 
and 
\begin{equation}
\label{eq:variance-ln-C}
\var (\ln \mathcal{C}) \propto \ell^{\mu} \quad \text{with} \quad \mu < 1 \,.
\end{equation}
Importantly, the characteristic magnitude of fluctuations of $\ln \mathcal{C}$, which is $\propto \ell^{\mu/2}$, is much smaller than its average value $\propto \ell$. Further, the number of terms in the sum  \eqref{eq:supp:GAB-C} scales only as a power law with $\ell$. The assosiated power-law factors  are subleading with respect to exponential factors. As a consequence, the power-law number of terms in Eq.~\eqref{eq:supp:GAB-C} results in subleading logarithmic corrections, see formulas below. 

For clarity of the argument, we make several assumptions, which simplify the analysis without affecting its essence. 
First, we assume that the distribution of $\ln \mathcal{C}$ is Gaussian. Second, in view of the exponential decay of $\mathcal{C}_{\boldsymbol{x}\boldsymbol{x}^\prime}$, we keep in the sum in Eq.~\eqref{eq:supp:GAB-C} only the terms with the minimal distance  $\boldsymbol{x} - \boldsymbol{x}^\prime = \ell$. The number of such terms is $4\ell$. Further, since $\mathcal{C}_{\boldsymbol{x}\boldsymbol{x}^\prime}$ in the localized regime is governed by trajectories that propagate nearly straight from $\boldsymbol{x}$ to $\boldsymbol{x}^\prime$, we assume that these $\sim \ell$ terms are statistically uncorrelated. Finally, in view of a rather broad distribution of $\ln \mathcal{C}$, we approximate the sum in Eq.~\eqref{eq:supp:GAB-C} by its largest term. The distribution of $P(\ln G_{AB})$ is then approximated by the extreme-value statistics applied to $\ln \mathcal{C}$; specifically, it corresponds to the distribution of the maximum among $\sim \ell$ independent samples drawn from the distribution of $\ln \mathcal{C}$.

Performing the analysis with the above approximations is straightforward, and we only quote the results. The average value of $\ln G_{AB}$ is shifted as compared to the average value of $\ln \mathcal{C}$ according to
\begin{equation}
\label{eq:average-ln-G}
\overline{\ln G_{AB}} - 
\overline{\ln \mathcal{C}}  \propto  \ell^{\mu/2}\ln^{1/2} \ell \,.
\end{equation}
Further, the variance of $\ln G_{AB}$ is reduced compared to the variance of $\ln \mathcal{C}$ by a logarithmic factor:
\begin{equation}
\label{eq:variance-ln-G}
\var (\ln G_{AB}) \sim
\frac{\var (\ln \mathcal{C})}{\ln \ell} \,.
\end{equation}
Clearly, the shift in Eq.~\eqref{eq:average-ln-G} is subleading with respect to the linear-in-$\ell$ term, Eq.~\eqref{eq:average-ln-C}, implying that $\ell_{\text{loc}}^{\text{(typ)}}$ is asymptotically the same for both observables. Similarly, the logarithmic factor in Eq.~\eqref{eq:variance-ln-G} is subleading with respect to the power-law factor in Eq.~\eqref{eq:variance-ln-C}, so that the exponent $\mu$ determining the asymptotic power-law scaling of the variance is the same for both quantities. 

While differences in average values and variances of $\ln \mathcal{C}$ and $\ln G_{AB}$ are subleading in the large-$\ell$ limit, they are sizeable for moderately large $\ell$. This explains differences between the two distributions (and, in particular, between the corresponding average values and variances) that are seen in Fig.~\ref{fig:PDF:C:localized}.

\end{document}